# When to Relax Social Distancing Measures? An ARIMA Based Forecasting Study

**Ramya Hariharan**[*]

**B.M.S. College of Engineering, Bangalore 560019**

*To whom correspondence should be addressed now at:

Department of Physics,

B.M.S. College of Engineering, Bull Temple Road,

Bangalore, 560019, India.

Tel.:+91-80-2660 3961 Ext: 2072; e-mail: ramyah.phy@bmsce.ac.in

**Abstract**

The spread of the novel coronavirus across various countries is wide and rapid. The number of confirmed cases and the reproduction number are some of the epidemiological parameters utilized in scientific studies for the analysis and prediction of the viral transmission. The positive rate, an indicator on the extent of testing the population, aids in understanding the severity of the infection in a given geographic location. The positive rate for selected countries has been considered in this study to construct ARIMA based statistical models. The goodness of fit of the models are verified by the investigation of residuals, Box-Luang test and the forecast error values. The positive rates forecasted by the ARIMA models are utilized to investigate the scope for implementation of relaxations in social distancing measures in some countries and the necessity to tighten the rules further in some other countries.

# 1. Introduction

In the first two decades of the 21$^{st}$ century, the re-emergence of infectious disease is on the rise. Predominantly originating from zoonotic viruses, the outbreak of viruses such as Ebola, Zika, H1N1 etc., have resurfaced frequently in the past few years [1,2]. Recently, severe acute respiratory syndrome coronavirus 2 (SARS-CoV-2), a novel virus which belongs to the family of human coronavirus has originated in the Wuhan province of China in December 2019 [3,4]. Through an intermediate host, the transmission of the virus from bat to humans has occurred. The virus has a higher infectivity rate as compared to the influenza viruses which is manifested by its high reproduction number. Owing to the severity of the respiratory illness World Health Organization (WHO) had declared SARS-CoV-2 as a pandemic in March 2020. The virus has rapidly spread across the globe and almost all the countries are engulfed under its net. As on 01 September 2020, as high as 213 countries around the world have reported a total of 21,095,532 confirmed cases of SARS-CoV-2. The global death toll has also reached about 757,779 deaths. All the affected countries are battling to reduce the transmission by proclaiming stringent guidelines on safety precautions, social distancing, lockdown, home/institutional quarantine and travel restrictions [5]. In spite of these measures, the number of confirmed cases in most of the countries is on the rise. In particular, highly populated countries such as Brazil, USA, India etc., have emerged as epicenters for the infectious disease [6,7]. Numerous research groups have utilized the epidemic data to understand and the trend and trajectory of the disease. Most of the studies have been conducted by analyzing the country-wise or city-wise data on the number of confirmed cases, recovery rates and mortality rates [8,9]. Machine learning, deep learning, artificial neural networks based algorithms have been utilized to forecast the transmission of the disease [10–12]. However, the re-emergence of SARS-CoV-2 cases in countries such as New Zealand, Spain, Germany, Iran etc., has raised speculation on the possibility of a second epidemic wave. The testing for infection which aids in the effective isolation of infected persons and also tracing their contacts is very important to overcome the viral dissemination [13]. Testing the population also helps in efficiently utilize the medical resources, which is being exploited in the pandemic situation.

In this study, an interesting epidemiological parameter, namely, the positive rate has been utilized to frame guidelines pertaining to social distancing measures. Positive rate is an indicator on the level of testing with respect to the extent of outbreak. The time series data collected from some of the worst affected countries of the world has been used to build the Auto-Regressive Integrated Moving Average (ARIMA) models. The countries are chosen so as to assess concerns such as increase in SARS-CoV-2 cases as a function of population, economy, testing rate and travel regulations. The study provides a new perspective to understand the current pandemic situation and also provides insight on efficiently using the available resources.

## 2. Materials and Methods

### 2.1 Data collection

The data on positive rate has been collected from open source database of Our World in Data [14]. As on the first week of September 2020, USA, Russia, South Africa, India, Mexico and Spain are some of the countries badly affected by the deadly virus. Therefore, the master dataset has been filtered to obtain the positive rates for the chosen countries. The data was collected from 1 April to 12 September 2020. The country-based positive rates are shown in Figure 1 which also highlights the countries selected in this study. The data was checked for missing values which was approximated by the corresponding monthly average.

## 2.2 Model building and performance metrics

The as collected data showed a time series behavior. In order to build a suitable model to forecast the trend in variation of positive rate, the steps such as test for stationarity, identification of parameters, estimation, evaluation of model performance and forecasting are performed on the collected data. Among the statistical models, the most powerful and robust procedure is the method established by Box and Jenkins [15]. The ARIMA model which is a combination of the autoregressive (AR) and moving average (MA) models has been used in this study to analyze and forecast the time series data [16]. The "Arima" function in the "forecast" package of R programming (version 3.4.2) was used to build the model. In addition, packages such as "timeseries", "Metrics" and "ggplot2" were used for forecasting, statistical analysis and visualization respectively.

## 3 Results and discussion

### 3.1 Building of country-specific ARIMA models

The positive rate values for the chosen countries such as USA, Russia, South Africa, India, Mexico and Spain are converted into their corresponding time-series plots. Depending on the nature of the time-series data, suitable ARIMA models are built for each of the countries. Initially, the unit root test is conducted to ascertain the non-stationarity of the time series. In order to apply the statistical theories, it is mandatory for the time series to be stationary. One of the popular tests to estimate the stationarity of data is the Augmented Dickey-Fuller (ADF) test [17]. It is observed from Table 1 that the data from all the chosen countries lacked stationarity as manifested by their P values > 0.05. The probability of significance, P-value, should be < 0.05 to confirm that the time series is stationary.

A non-seasonal ARIMA model is generally represented by the parameters (p,d,q). The primary step in fitting an ARIMA model is the determination of the order of differencing, 'd' necessary to stationarize the series [18]. The country-wise estimated order of difference required to yield the positivity data series stationary are shown in Table 1. The ADF test has been repeated on the differenced data series which showed P-values <0.01 thereby confirming its stationarity. The parameter 'p' which captures the autoregressive (AR) nature of ARIMA is estimated from the Partial Auto Correlation (PACF) function. PACF provides information on the relationship between an observations in a time series with observations at prior time steps with the relationships of intervening observations removed. The appropriate values of the lag order (p) for the various ARIMA models are estimated from the PACF correlogram as tabulated in Table 1. The values of the order of moving average (q) are obtained by identifying the most appropriate lags for moving average (MA). The q-values estimated by identifying the autocorrelation (ACF) function that exceed the confidence boundary lag in the autocorrelation plot. The values of q for the different ARIMA models corresponding to the chosen countries are shown in Table 1. Subsequently, the ARIMA models are built using the least-squares estimation process.

### 3.2 Validation of the ARIMA models

The accuracy of the ARIMA models is diagnozed using Akaike's Information Criterion (AIC) and the Schwartz Bayesian Information Criterion (BIC). AIC, shown in Equation 1, is a widely used measure of a statistical model to quantify its goodness of fit and parsimony. A good model is identified as the one which has minimum AIC among all the other models [19]. The models are selected based on this criterion and the corresponding AIC values are listed in Table 1. Similarly the BIC, as shown in Equation 2 is another criterion for model selection which estimates the trade-off between model fit and complexity of the model. The models with lower BIC value indicates a better fit with the actual values as tabulated in Table 1.

$$AIC = -2 \times \ln(L) + 2 \times k \qquad (1)$$

$$BIC = -2 \times \ln(L) + 2 \times \ln(N) \times k \qquad (2)$$

Where L is the likelihood value, N is the number of measurements recorded and k is the number of estimated parameters.

### 3.3 Investigation on the residuals

The validity of the ARIMA models are investigated by analyzing the residual values. Residuals are one of the key factors which provide information on the model's ability to learn from the time series data which in turn controls the accuracy of the forecast. The residuals-time plot for the ARIMA models of the countries chosen in this study are shown in Figure 2. Excepting Figure 2f, the other residual-time plots confirm low residual values of about -3 to 3. In Figure 2f, the higher residual values may be attributed to the presence of one or two outlier values. However, the outliers are also retained in this study to build a realistic model.

Similarly, the estimated autocorrelation coefficients (ACF) of the residuals pertaining to the various models are shown in Figure 3. It is evident that for all the models, the lags shown in Figure 3 occur well within the confidence interval. It is also noticed from Table 2 that the ACF are statistically insignificant which implies that the residuals have random values and confirm the lack of noticeable correlation in the residuals series [20].

The histogram of residuals are shown in Figure 4 corresponding to the chosen countries. The histograms of all the ARIMA models show a predominant normal distribution trend of the residuals. They also confirms the lack of significant variance. The mean values are also noted to be near-zero. The observation is also confirmed from the low mean error (ME) values compiled in Table 2. The investigation on the residuals confirm that the built ARIMA models have a good fit with the actual values.

### 3.4 The Box-Ljung test

The ARIMA models are also verified by the Box-Ljung test which provides a statistical evidence of a good fit [18]. The recorded P values for all the ARIMA models are tabulated in

Table 1. All the values are noted to be higher than 0.05 which confirms that the forecasted values are fitted well with the actual values. The performance of the ARIMA models are further vouched by the low root mean square error (RMSE) and low mean absolute error (MAE) values as summarized in Table 2.

### 3.5 Forecasting using ARIMA models

Figure 5 shows a country-wise 30-day forecasted values of positive rate with a confidence interval of 80%. Figure 5a shows a steady reduction in the positive rate for USA from 6% to 4.8% in the forthcoming month. The forecast confirms that the current testing rate in USA is adequate to isolate the infected population. WHO has provided guidelines to observe the positive rate for 14 days before taking decision on relaxing social distancing measures. It has also recommended a positive rate of 5% or lesser as a metric to evaluate the viral spread and to frame government policies [21]. Based on the forecasted positive rate, in the forthcoming months USA would comfortably meet the criterion and policies for social distancing shall be taken favorable to allow people interaction and movement.

Figure 5b shows the forecasted trend of positive rate in Russia. The vast country with a large population is one of the worst affected nations by the SARS-CoV-2 pandemic. The forecasted values show a gradual increase of positive rate from 1.69% to 1.92%. The increase may be attributed to the shift in weather to colder season in the northern hemisphere. Colder seasons offer favorable conditions for the survival of the pathogen. However, since the values are much lesser than 5% and consistent, the current testing rates are adequate and support the decision to relax the rules enforced due to the pandemic. In a similar manner, one of the worst affected African countries, South Africa, shows a steep decrement in the forecasted positive rate value of 11.69 % to 0.1 % as noticed from Figure 5c. The forecasted values indicate that the testing rates are adequate as confirmed by Figure 6a and reveals the potential towards relaxation of social distancing rules in the near future. The prediction is also supported by Figure 6b which shows that the country adopts open public testing policy and is not selective about the conduction of the test. The forecasted positive values for India, has shown a marginal decrease from 7.71 % to 7.12% as observed from Figure 5d. India is a highly populated country and also has been consistently recording the highest number of daily confirmed cases in the past few weeks. The forecasted values clearly indicate that the current testing rate is insufficient as confirmed by Figure 6a. The number of tests should be increased by at least 20-30% times to reduce the positive rate below the WHO prescribed threshold. Figure 5e shows an alarmingly high forecasted positive value of about 50% for Mexico. The high positive rate indicates that the government is very selective about conduction of the tests as noted from Figure 6b. It appears as though, the patients who seek medical attention are only being considered for testing for infection. However, an asymptotic patient could be responsible for the silent spread of the pathogen to a large volume of population. In order to reduce the viral transmission, the government should be proactive and increase the testing rate by at least 60-70%.

Figure 5f shows the forecasted positive rates for Spain. The values show a moderate increment from 9.09% to 15.14%. The positive rate is higher than the WHO prescribed limits and the country should increase the testing rate by at least 30-40% to reduce the positive rate to below 5%. In the case of Spain this measure is very critical because it is a European Union country which is well-connected by land and other means of transport with most of the European countries. Hence the government should not only increase the testing rate, but also implement stringent rules until the positive rate falls below 5%.

This study provides an insight on the current and forecasted trend in the positive rate in the selected countries. The countries such as USA and South Africa, similar to Australia and South Korea, are on the path of attaining low positive rates [22]. Hence, the guidelines introduced to mitigate the transmission of SARS-CoV-2 such as restriction on people movement and social distancing could be relaxed in a month in these countries. However, countries such as India, Mexico and Spain should adopt more precautions to protect their population by increasing the testing rates. The social distancing rules should also be strictly enforced in these countries for at least a few more months until the positive rate is <5%. Countries such as Russia should be more proactive in maintaining the low positive rates in order to avoid the outbreak of a second wave of the pandemic. This knowledge is important to properly assess the key decision to be implemented on social distancing and lockdown policies. These measures are essential because the SARS-CoV-2 pandemic holds the potential to spread more widely and quickly which can have a devastating impact on not only the economy of the affected country but also the global economy.

**4. Conclusion**

In this study, the positive rate for different countries is utilized to build ARIMA based forecasting models. The model has been carefully built by testing for stationarity, appropriately choosing the key parameters and validating its accuracy. The verification of the models is conducted by observing the residuals and from the results of the Box-Ljung test. The forecasted values for USA and South Africa showed than in 30 days, the positive rate shall be consistently below 5%. This would pave way to relax the existing stringent measures adopted by the government to reduce the viral transmission. However, in countries such as India, Mexico and Spain, the positive rate is beyond the safe limit of about 5%. In India, even though the positive rate is not inherently high, precarious measures such as active testing would prevent a steep increase in the positive rate. In Mexico, the current testing rates are highly insufficient which need to be increased by at least 60-70%. In Spain, the testing rate should be consciously increased by 30-40% in order to prevent the cross-country spread of the pathogen.

**Funding**

This research did not receive any specific grant from funding agencies in the public, commercial, or not-for-profit sectors.

**CRediT authorship contribution statement**

**Ramya Hariharan**: Conceptualization; Data curation; Methodology; Resources; Software; Validation; Visualization; Writing - original draft, review & editing.

**Declaration of conflicting interest**

The author declares that they have no conflict of interest.

**Acknowledgements**

The author is thankful to Our World in Data organization for the valuable data. The author profoundly thanks the Management and the Principal of B.M.S. College of Engineering, Bangalore for their support.

Pathological findings of COVID-19 associated with acute respiratory distress syndrome, Lancet Respir. Med. 8 (2020) 420–422. https://doi.org/https://doi.org/10.1016/S2213-2600(20)30076-X.

Table 1 The recorded ARIMA model building and validation parameters as a function of the chosen countries

| Countries | P (ADF) | p | d | q | AIC | BIC | Box-LJUNG |
|---|---|---|---|---|---|---|---|
| USA | 0.424 | 7 | 2 | 7 | 150.6 | 196.8 | 0.720 |
| Russia | 0.055 | 6 | 2 | 5 | -261.4 | -224.5 | 0.061 |
| South Africa | 0.971 | 6 | 2 | 5 | 249.8 | 286.6 | 0.963 |
| India | 0.947 | 7 | 2 | 4 | 077.2 | 114.1 | 0.351 |
| Mexico | 0.878 | 5 | 1 | 6 | 454.3 | 490.9 | 0.898 |
| Spain | 0.744 | 2 | 2 | 5 | 329.0 | 351.8 | 0.489 |

Table 2 The statistical performance metrics for the ARIMA models built as a function of the selected countries

| Countries | ME | RMSE | MAE | ACF |
|---|---|---|---|---|
| USA | -0.012 | 0.330 | 0.213 | -0.016 |
| Russia | -0.008 | 0.097 | 0.070 | 0.009 |
| South Africa | -0.027 | 0.471 | 0.331 | 0.004 |
| India | -0.01 | 0.276 | 0.180 | 0.017 |
| Mexico | 0.142 | 0.929 | 0.726 | -0.008 |
| Spain | 0.046 | 0.797 | 0.366 | -0.067 |

**Figure Captions**

**Figure 1**: Trend in Positive Rates for the countries in the world as on September 2020 (Source: Official data collected by Our World in Data [14])

**Figure 2**: Residual versus Time plots for the ARIMA models for (a) USA (b) Russia (c) South Africa (d) India (e) Mexico (f) Spain

**Figure 3**: Autocorrelation correlogram for (a) USA (b) Russia (c) South Africa (d) India (e) Mexico (f) Spain

**Figure 4**: Histogram of residuals of the ARIMA models corresponding to (a) USA (b) Russia (c) South Africa (d) India (e) Mexico (f) Spain

**Figure 5:** The ARIMA model forecasted positive rates for (a) USA (b) Russia (c) South Africa (d) India (e) Mexico (f) Spain

**Figure 6:** The global trend in (a) The number of tests per 1000 persons conducted (b) The open testing policy (Source: Official data collected by Our World in Data [14])

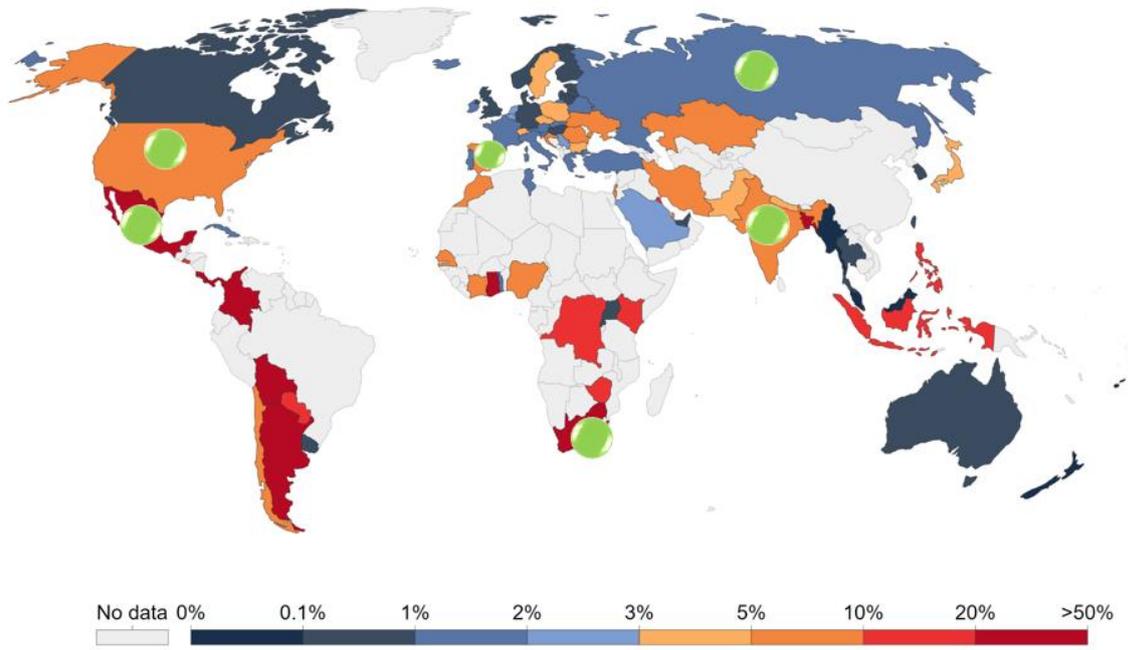

**Figure 1 R Hariharan**

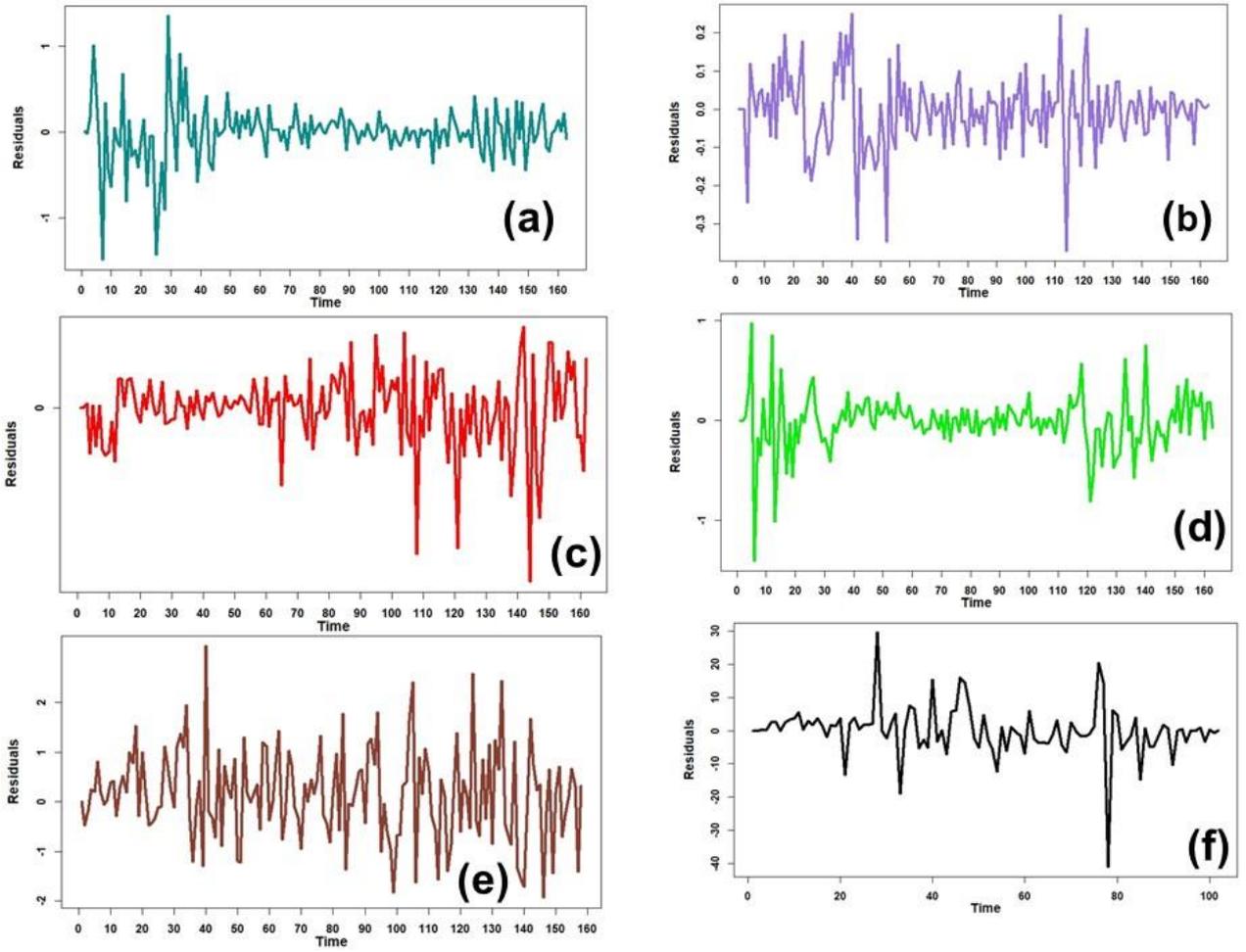

**Figure 2 R Hariharan**

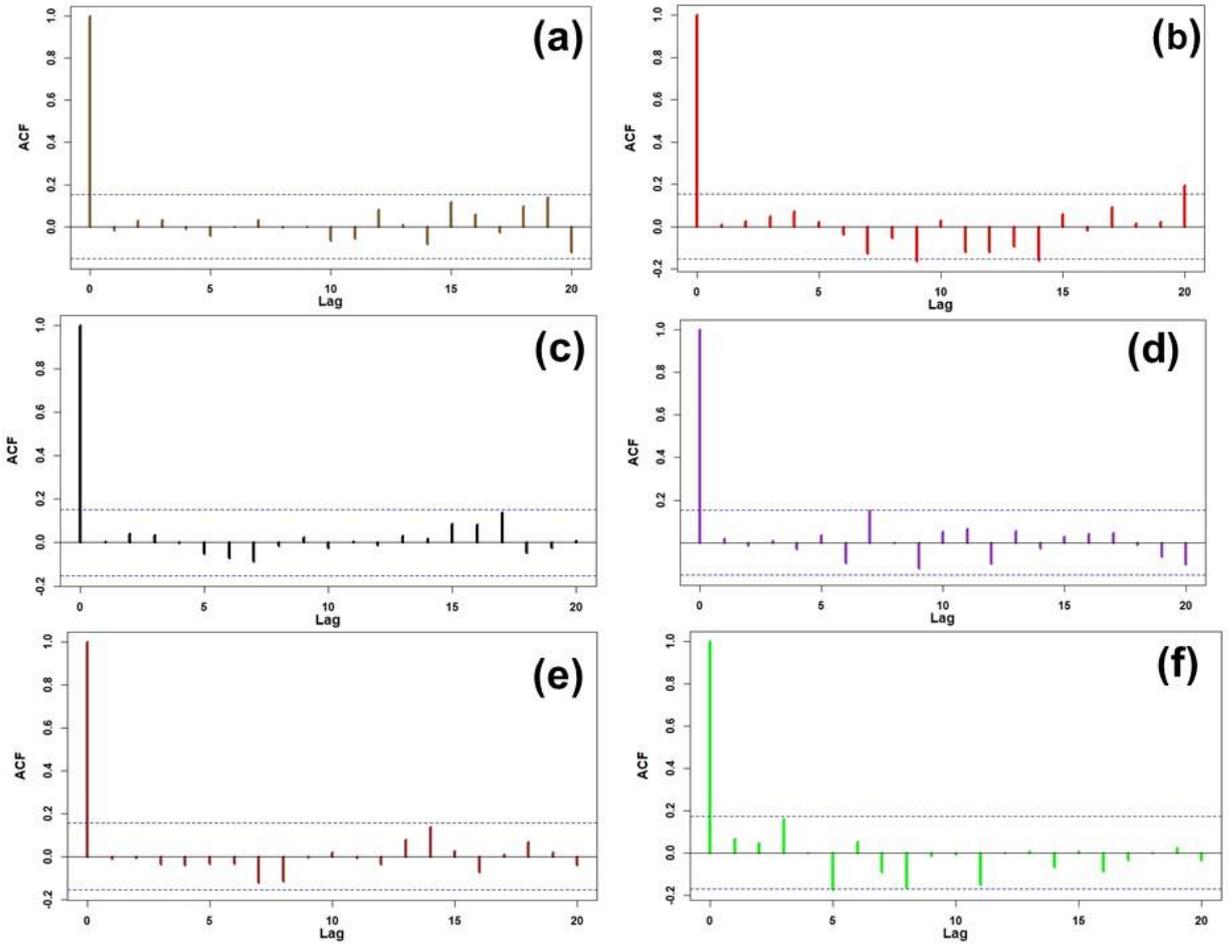

**Figure 3 R Hariharan**

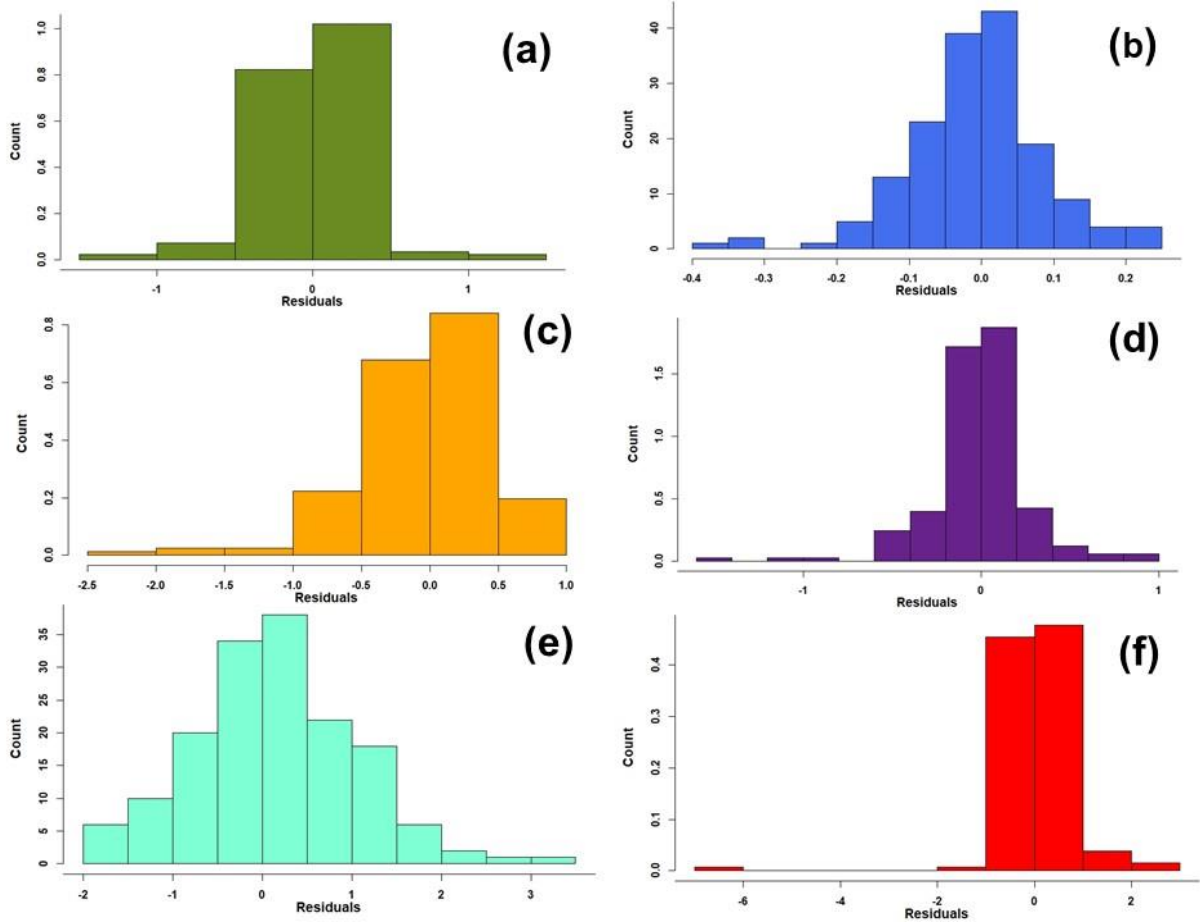

**Figure 4 R Hariharan**

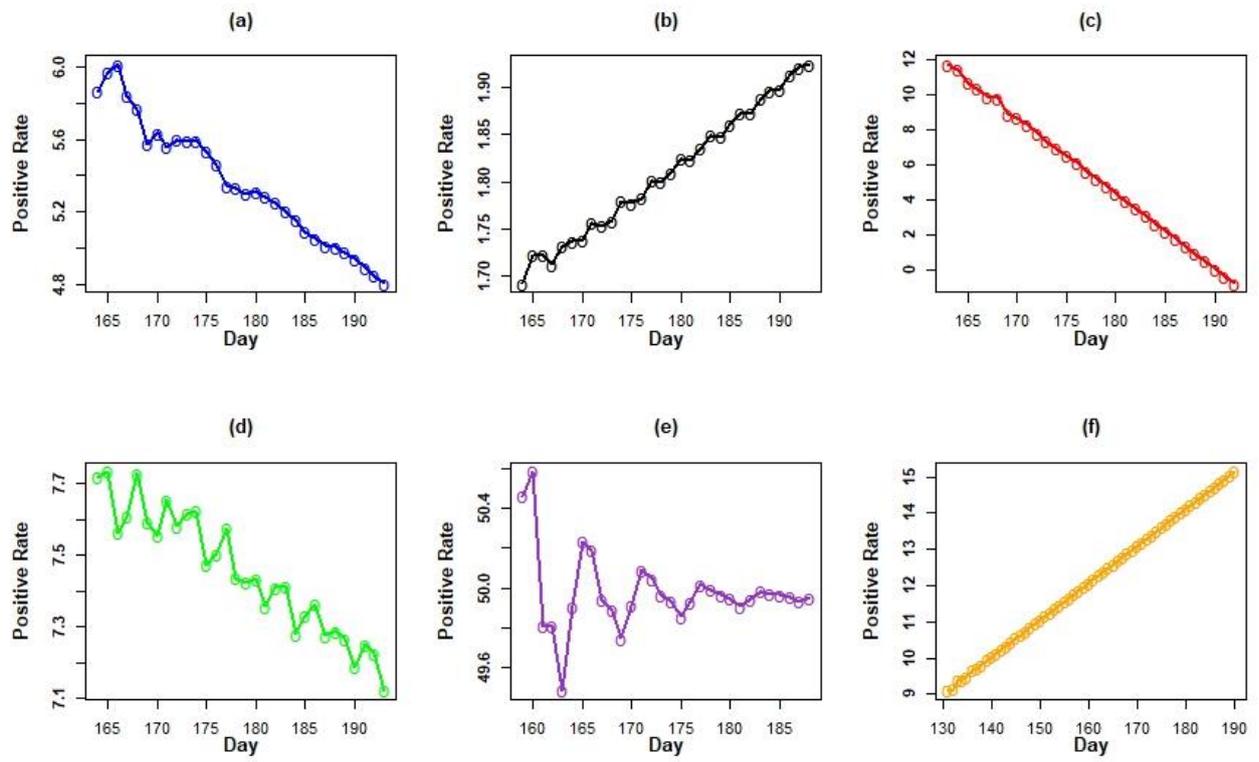

**Figure 5 R Hariharan**

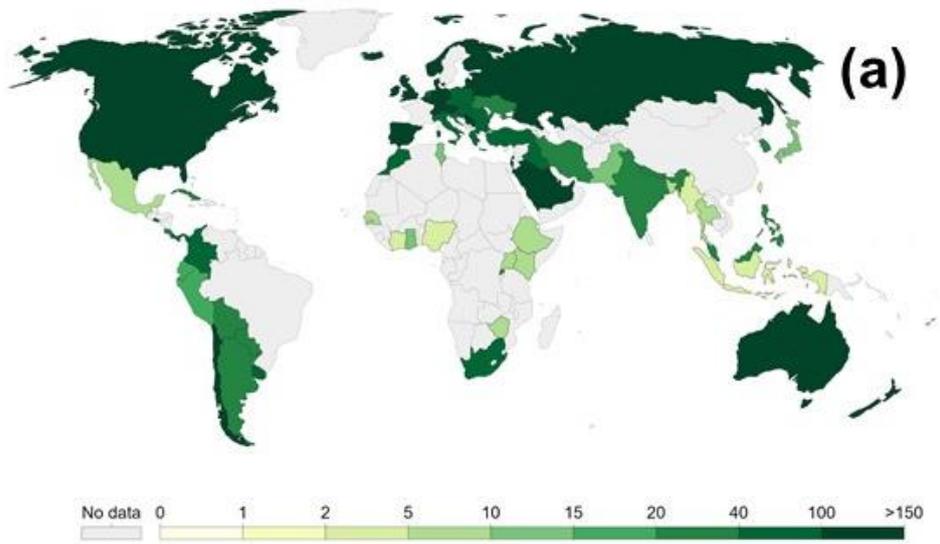
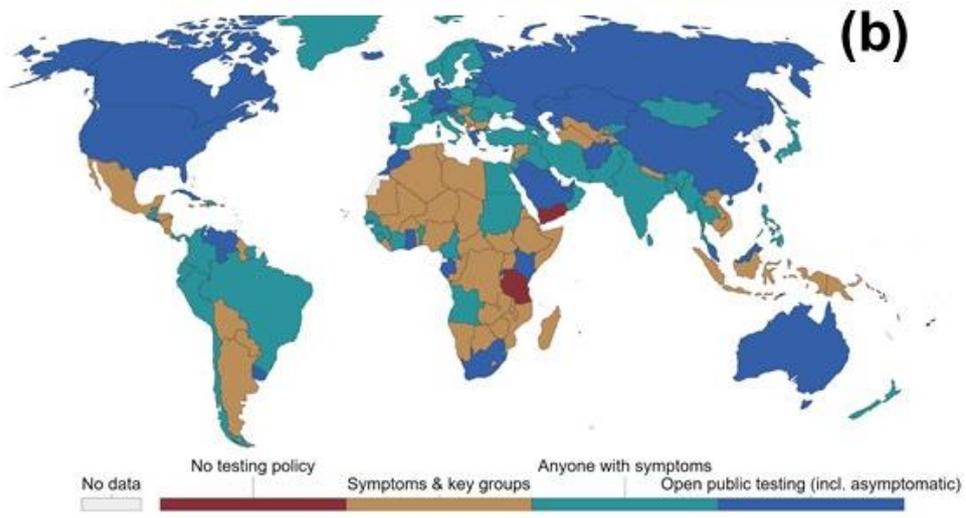

**Figure 6 R Hariharan**